# Electrotunable friction with ionic liquid lubricants: how important is the molecular structure of the ions?

*Oscar Y. Fajardo[1*], Fernando Bresme[2*], Alexei A. Kornyshev[2*], and Michael Urbakh[1*]*

[1]School of Chemistry, Tel Aviv University, 69978 Tel Aviv, Israel

[2]Deparment of Chemistry, Imperial College London, SW7 2AZ London, U.K.

Using non-equilibrium molecular dynamics simulations and a coarse grained model of ionic liquids, we have investigated the impact that the shape and the intramolecular charge distribution of the ions have on the electrotuneable friction with ionic-liquid nanoscale films. We show that the electric-field induces significant structural changes in the film, leading to dramatic modifications of the friction force. Comparison of the present work with previous studies using different models of ionic liquids indicate that the phenomenology presented here applies to a wide range of ionic liquids. In particular, the electric-field-induced shift of the slippage plane from the solid-liquid interface to the interior of the film and the non-monotonic variation of the friction force are common features of ionic lubricants under strong confinement. We also demonstrate that the molecular structure of the ions plays an important role in determining the electrostriction and electroswelling of the confined film, hence showing the importance of ion-specific effects in electrotuneable friction.

Corresponding author: urbakh@post.tau.ac.il

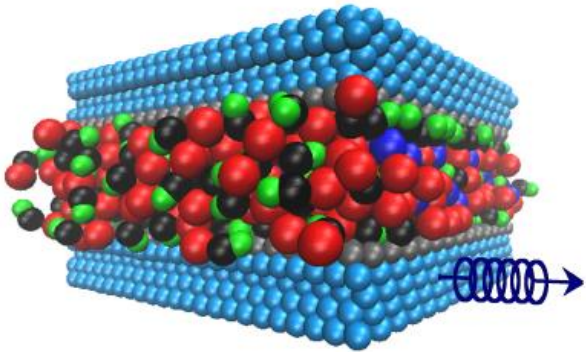
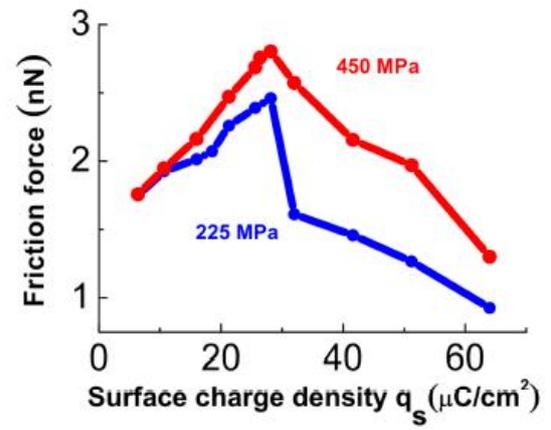
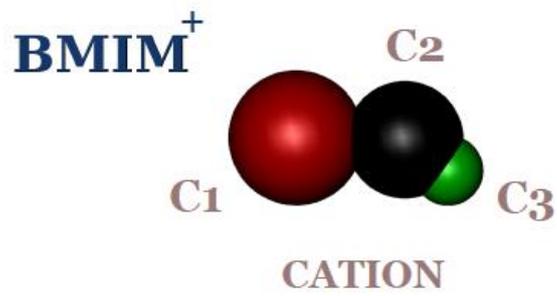
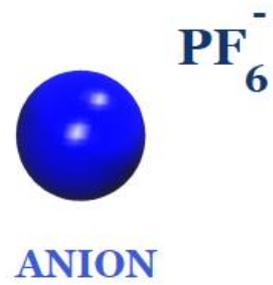

Nanotribology is a rapidly growing area in physics, chemistry and engineering [1,2], of interest both from the application and fundamental points of view. The study of the nanoscale aspects of lubrication is attracting particular attention, as under normal loads, the active lubricant is just a few molecular layers thick. Hence, a molecular approach is needed to understand the structure of the confined liquids and their response to the load, shear, temperature and applied electric field. This is instrumental for a rational design of industrial fluids with the desired and tunable lubricity.

The development of new promising lubricants requires to identify the generic, as well as the specific features of the fluids under nanoscale confinement. **R**oom **t**emperature **i**onic **l**iquids (RTILs) have emerged as a new and exciting class of lubricants.[3] They attract strong attention because of their unique physical properties, in particular their low vapor pressure and the virtually unlimited variety of RTILs and their mixtures.[4,5] Pure RTILs are in general solvent-free electrolytes, although they are hygroscopic and it is difficult to eliminate completely the water. [6] RTILs can be manipulated using electrical potentials, an approach that is particularly useful under confinement, since the confining surfaces can be charged to build up an electric field across the nanoscale film, inducing strong structural changes in the film [7,8]. The response of RTILs to external fields is not modified significantly when small amounts of organic solvent are added to reduce their viscosity. [9]

Experimental studies of friction with nanoscale RTILs have uncovered the phenomenon of *quantized friction*, *i.e.* the dependence of the friction coefficient on the discrete number of confined ionic layers [10], as well as demonstrated how electric fields influence lubricity [11,12]. So far, the latter has been studied only with the friction force microscope (FFM), which due to the limited resolution allows mainly measurements with monolayer films confined between the tip and the surface.[11,12,13]

There have been a number of reports devoted to molecular dynamics simulations of the nanoscale friction with model ionic liquids [14,15,16]. The first theoretical investigation of electrotuneable lubricity in RTILs was reported recently by us [8]. That study has shown dramatic changes of the structure of lubricating film in normal and lateral directions with the increase of the surface charge. We found that the latter induces stronger binding of the first layer of counterions to the substrate, resulting in a more ordered structure ('solidification') of the first layer; at small charges due to the overscreening effect [7] the second layer is yet strongly bound with the first one. With further increase of surface charge, the interaction with the second layer gets weaker when overscreening disappears, and the slippage plane shifts away from the solid-liquid interface to the interior of the film. At intermediate surface charges, the friction force reaches a maximum and subsequently decreases with increasing surface charge. We suggested that this mechanism may be used to control friction, simultaneously enabling the protection of the surface against wear, since the first adsorbed layer of ions is strongly adsorbed on the substrate.

These important predictions were obtained using a 'toy-model' of RTILs, where the cation and anion are modeled using charged Lennard-Jones spheres of dissimilar size. However, the complex shape of RTIL ions, as well as their charge distribution, may play a role in determining the structure of the nanoconfined film. One question we wish to tackle in this paper is: how does the RTILs molecular structure influence the friction mechanism, and in particular the electrotunable friction response? To address this question we examine a more realistic model of RTIls, taking advantage of the molecular coarse-grained approach. We use a model that has been parameterized to describe [BMIM][PF6].[17] We will demonstrate that this model recovers the general electrotunable behavior predicted on the basis of the spherical ion model. Further, the new model highlights the importance of the chemical specificity of the ions, in particular regarding electrostriction and electroswelling phenomena. We thus uncouple the physical and chemical factors determining the microscopic mechanisms of electrotunable friction.

*Model*. The ionic liquid was modeled using a coarse grained representation, where the cation is represented by three beads restrained with rigid bonds and angular terms, while the cation is modeled as a single sphere (see Fig.1) This force-field for MD simulation has been proposed in Ref. [17] to model 1-butyl-3methylimidazolium hexafluoro phosphate [BMIM]$^+$ [PF$_6$]$^-$. To emulate the structure of mica surfaces used in the surface force apparatus (SFA) experiments, [10] the fcc {111} plane was selected as the friction plane in contact with the ionic liquid. To study electric-field effects, partial charges were assigned to each atom lying on the substrate surfaces that confine the ionic liquid. The surface charge density, $q_s$, was varied between $0 < q_s < 60$ uC/cm$^2$. The typical values of charge density at the mica-RTILs interfaces studied in SFA experiments lie in this interval.[10] In addition to the Coulombic interactions, the beads interact with other beads or atoms in the solid plates through short range Lennard-Jones interactions with the parameters described in the *Methods* section.

We simulated the friction process by pulling one of the solid plates through a spring of stiffness, K$_{dr}$ =16 N/m at a constant velocity, typically $V_{dr}$=10 m/s, while the other plate was kept in the original position by using a restraining harmonic potential. The mobile solid plate was subjected to a normal force, equivalent to a normal pressure in the range of 225-450 MPa, in order to emulate experimental conditions. In this range of pressures the confining region contains two or three layers of cations. The simulation set up discussed above leads to stick-slip motion compatible with experimental measurements of friction.[10]

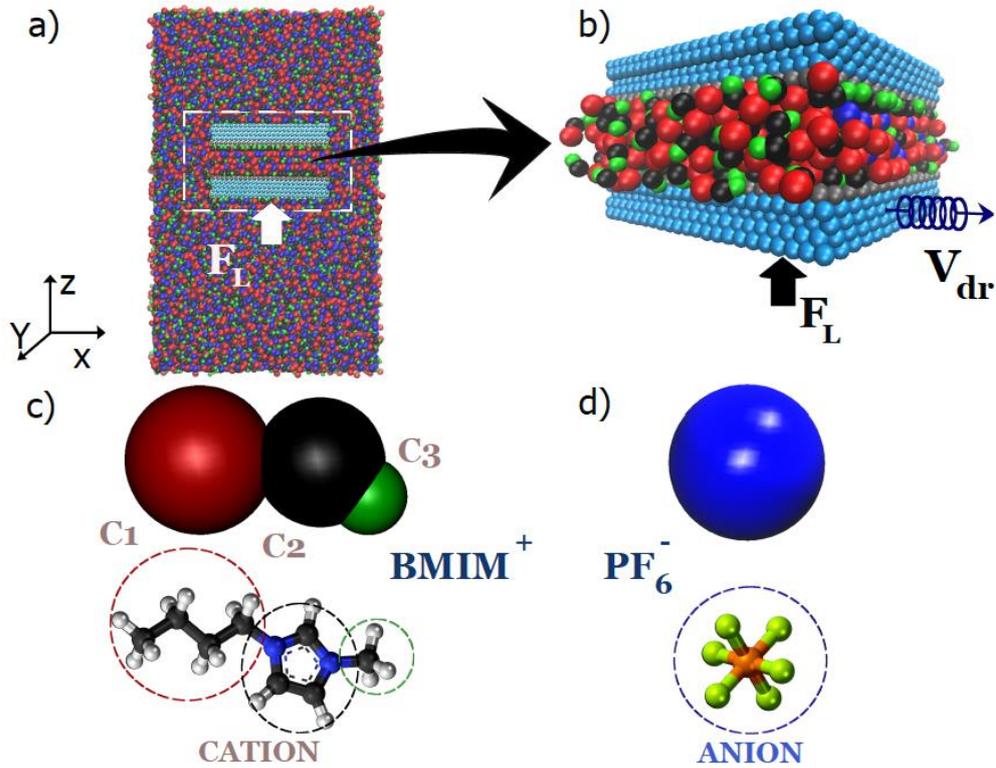

Fig.1 (a) Sketch of a typical simulation set up with the solid plate in a fully periodic simulation containing the ionic liquid. (b) Snapshot showing the confined ionic liquid under a normal load, $F_L$ acting on the mobile plate, which moves along the y axis at constant velocity $V_{dr}$. The simulation corresponds to the surface charge density, $q_s$=16 µC/cm$^2$, on the solid plates. The charged atomic layers used for the electrotuneable friction are highlighted in a gray color. The ionic liquid consists of BMIM$^+$ (c) cations and PF$_6^-$ (d) anions modeled using a coarse grained force field.

All the computations presented below correspond to the strength of ion-substrate interaction, $\varepsilon_{is} = 22.5$ kJ/mol, which leads to a full wetting state when the ionic liquid is in contact with the model surface (see Supporting Information (SI)). We have also performed simulations using weaker interactions, $\varepsilon_{is} = 1.25$ kJ/mol, resulting in the partial wetting at zero surface charge, but still full wetting at surface charge density, $q_s$>8µC/cm$^2$. The change in the ion-substrate interaction has some impact on the magnitude of the friction force, but it does not modify the dependence of the friction force on the surface charge and the microscopic mechanism of friction, hence we focus our discussion on the $\varepsilon_{is} = 22.5$ kJ/mol system. The results obtained for the lower value of $\varepsilon_{is} = 1.25$ kJ/mol are presented in the SI.

*Nonmonotonic dependence of the friction force and the shift of the slippage plane with surface charge variation.* So far the SFA experiments have not considered the variation of the surface charge, although a work in this direction is in progress with a new generation of SFA with solid substrates covered by graphene films [18]. In the FFM experiments the charge variation was implicit, but in that case the voltage between the surfaces was the control variable, rather than the charge density [11,12]. The surface charge adjusts the voltage self-consistently, subject to the electrical capacitance of the frictional contact, which itself may change during sliding. An approach that considers the system response at different fixed values of the charge allows, nevertheless, the investigation of the intrinsic mechanism of the electric field effect on friction. We exploit this notion in the present paper.

(a)

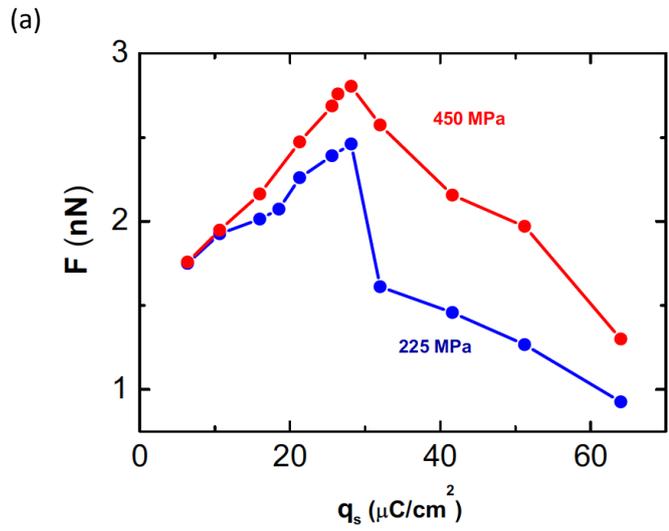

(b)

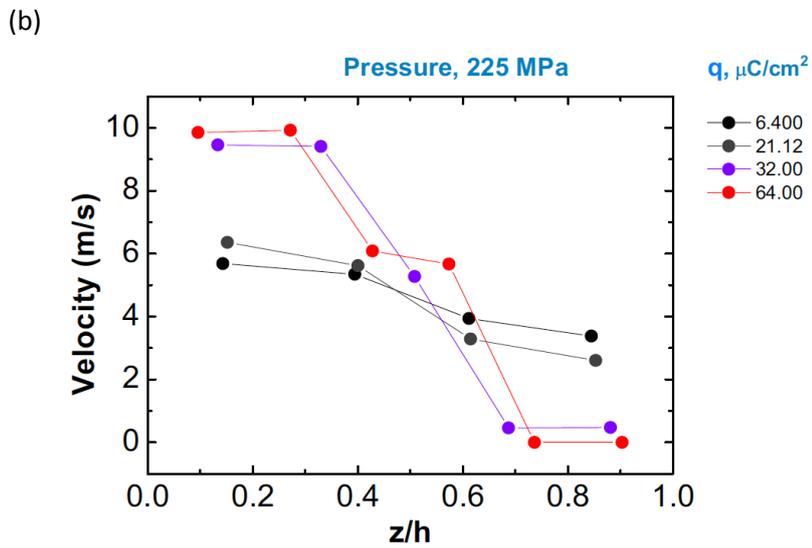

(c)

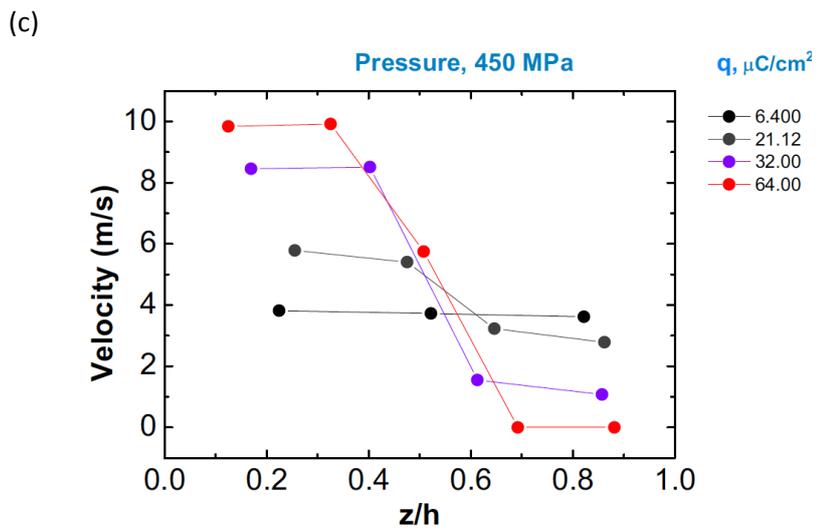

**Fig.2.** Electrotuning friction through the shift of the slippage plane. (a) Friction force as a function of surface charge density as calculated for two values of normal pressure. (b,c) Moving the lower solid plate keeping fixed the upper plate gives rise to a velocity profile across the RTIL-film; representative profiles are shown for different values of surface charge density at a fixed normal pressure. Note that for each surface charge density the distance between the plates is different; the normal coordinate *z* is scaled by the corresponding distance.

Figure 2a shows that the simulated friction force as a function of the surface charge density adopts a 'volcano' shape. The maximum friction corresponds to the charge at which the slippage plane shifts from the solid-liquid interface to the interface between the first, predominantly cation layer and the second, mostly anion layer. Indeed, for both normal pressures Figs.2b and 2c show a dramatic change of the velocity profile near this point.

The value of charge in typical surface force apparatus experiments with RTILs confined between mica is believed to be close to 32 μC/cm². At this value, we find the slippage plane lying in the middle of the film, which corroborates the assumptions used for interpretation of the experiments reported in Refs. [19].

The predicted volcano shape of the friction force and the shift of the slippage plane are independent of the parameters employed for the Lennard-Jones interaction between the cations and the substrates. We have checked this point systematically for a wide range of parameters and the corresponding results are presented in the SI.

Note that the choice of ion-solid Lennard-Jones interactions has a measurable impact on the contact angles of the ionic liquid with the solid substrate. For neutral surfaces we found that the contact angles are $67^0$ for $\varepsilon_{is} = 1.25$ kJ/mol, whereas complete wetting is observed for $\varepsilon_{is} = 22.5$ kJ/mol (see SI and reference [20] for details on the computation of contact angles). Increasing the surface charge results in a wetting state for all tested values of ion-solid interactions. This observation is compatible with experimental data that reported either full wetting or very low contact angles for RTILs–mica interfaces [21,22].

*Reorientation of cations and the RTIL-film thickness.* There are two preferential orientations for the cations adsorbed at a negatively charged surface. One of them is flat, the other one tilted, with the site with lower charge (C1) pointing into the liquid (c.f. Fig. 3a). The population of these two states changes with the surface charge density (c.f. Fig. 3b). For low $q_s$, the two states are almost equally populated. With an initial increase of the negative charge, anions of RTIL leave the surfaces, and the fraction of cations in the flat orientation increases. With further increase of the charge, more cations would energetically benefit from adhering to the surfaces, and to be accommodated there, they have to tilt. This effect is similar to that observed in classical electrochemistry, where the orientation of dipolar solvent molecules takes place upon varying the electrode polarization [23,24]. The charge-induced reorientation of the cations has a profound effect on friction, since the increase of the fraction of tilted cations leads to a reduction of the electrostatic attraction between the cation and the anion layers, hence facilitating the shift of the slippage plane from the solid-cation interface to the interface between the contacting cation and anion layers. In FFM experiments, where the electric field acting on the cations adsorbed on the electrode surface, and correspondingly their orientation, change when the tip approaches them, the reorientation of cations may lead to an additional route for energy dissipation, thus strongly influencing the potential dependency of friction. [25]

The electric field induced reorientation of cations might be difficult to measure directly, at least in existing SFA. However such reorientations would inevitably cause variation of the film thickness. The more tilted is the orientation, the thicker the film. In particular, the flattening of the orientation at low charges (see Fig.3b) is followed by the dip in the film thickness observed in Fig.3c. At large surface charge density the vast majority of cations are tilted, maximizing the thickness of the film.

(a)
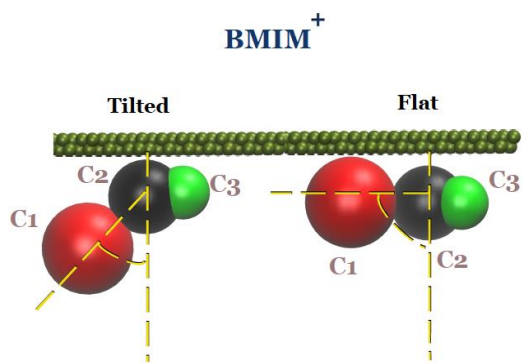

(b)
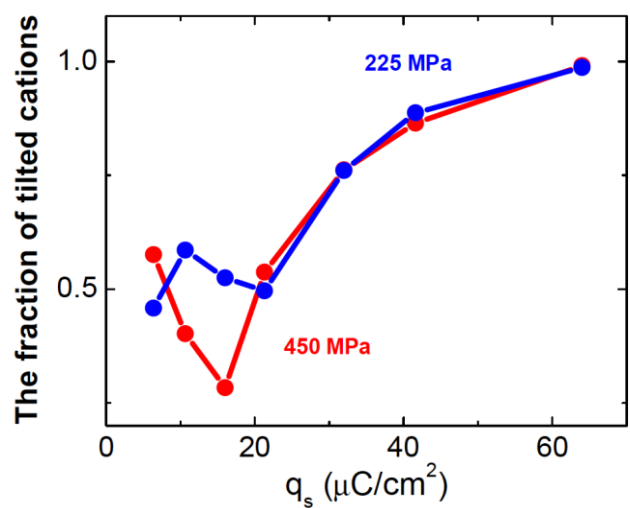

(c)
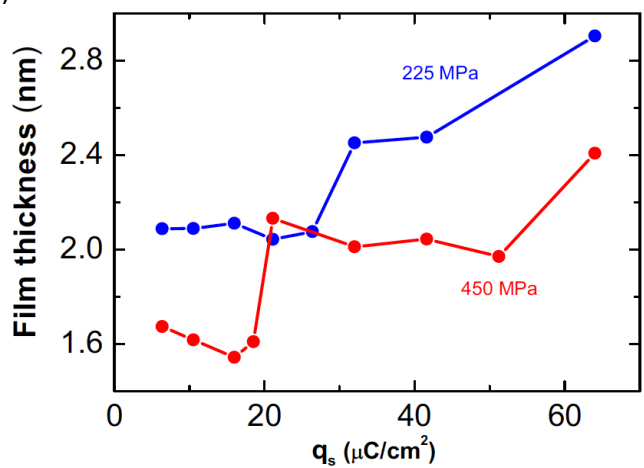

**Fig.3.** At a constant normal pressure, the charge induced reorientation of the cations affects the RTIL film thickness. (a) Sketch of the two preferential orientations of cations in the layers adjacent to the surfaces: 'flat' and 'tilted'. Dependence of the fraction of tilted cations in the first layer (b) and of the thickness of the RTIL film (c) as a function of surface charge density, shown for two normal pressures.

*Charge induced structural changes in the film.* In Table 1 we gather some structural characteristics of the studied nanoscale RTIL film, pertinent for understanding the electrovariable friction. They table shows how the film responds to the increase in negative charge on the surfaces.

**Table 1.** Charge-induced modification of the structure of the nanoscale RTIL film at normal pressure of 450 MPa.

| $q_s$ µC/cm$^2$ | Across-the-film number of beads representing **anions** individual compartments of cations - **C1**,**C2**,**C3**, as in Fig.1 | Structure factor (in reciprocal space) for the distribution of the **C3** groups in the layers adjacent to the surfaces. | Side view snapshots showing the positions of **anions** and of the **C1**,**C2**,**C3** groups of cations confined between the solid plates |
|---|---|---|---|
| -6.4 | 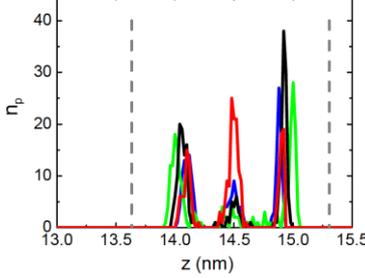 | 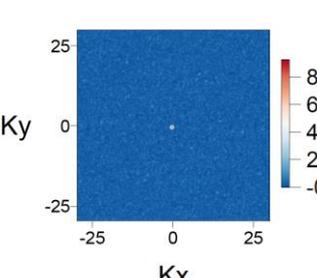 | 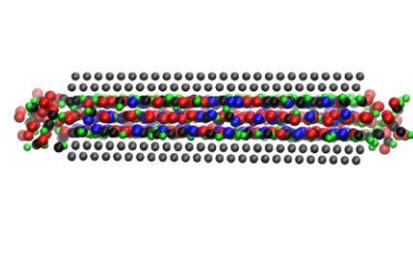 |
| -16 | 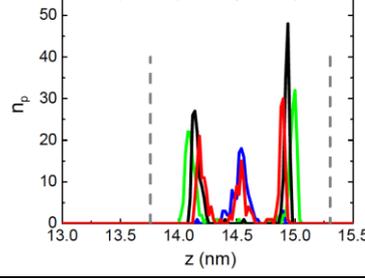 | 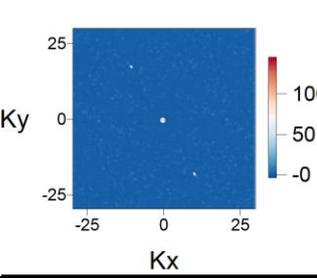 | 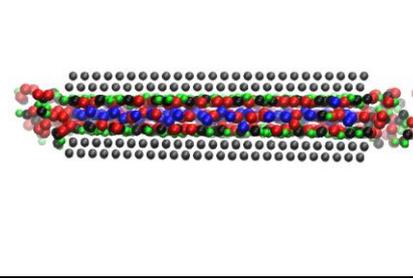 |

| -32 | 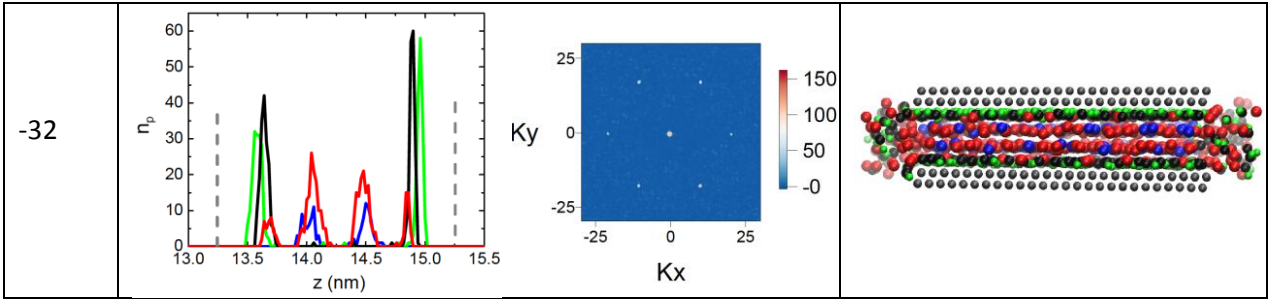 |

The first column in Table 1 gives the value of surface charges investigated (more charge points are available in SI). The panels in the second column represent the distribution of different charged groups across the film. The vertical dashed lines show the position of surface layers of substrate atoms. First, when the charge is small the film is liquid-like and the anions are evenly distributed across the film. As the negative charge increases, the anions deplete from the solid/liquid interface and form a well-defined layer between the two cation layers; for surface charges between 10 and 20 $\mu C/cm^2$ the three layer structure is clearly seen (c.f. figures in columns 2 and 4 for $-16\mu C/cm^2$). With further charging the population of tilted cations increases and the anions get less localized in the middle plane, penetrating in between the C1 groups of the tilted cations; this is accompanied by the increase of the film thickness. For the highest surface charge the friction reaches its maximum. This is connected to the shift of the slippage plane from the solid-liquid interface to the interface between the cation- and anion-rich layers.

The column 3 in Table 1 displays the lateral structure factors for the C3 groups in the cations in the layer adjacent to the surface. The structure factor is defined as $S(\mathbf{k}) = \frac{1}{N}\left|\sum_{j}^{N} e^{-i\mathbf{k}\mathbf{r}_j}\right|^2$, where $\mathbf{r}_j$ are the coordinates of the $j$ C3 group, and $N$ is the total number of C3 beads; $\mathbf{k} = (k_x, k_y)$ is a wave vector in the plane of the layer. For small charges the structure is liquid like (we see only one central peak of $S(\mathbf{k})$ at $k_x = k_y = 0$). With further increase of the negative charge, new peaks in $S(\mathbf{k})$ arise, which correlate with the periodicity of the negative charge pattern of the underlying surface. The intensity of these peaks grows with further charging, which indicates that the structure of the first layer becomes solid like.

Column 4 in Table 1 displays the simulation snapshots of the confined RTILs, which visualize the above mentioned trends, namely – the exit of anions from the first layer, the formation of the well-defined layer of anions localized between the cation layers, the shift of the C1 groups away from the solid-liquid interface, and the increase of the film thickness.

*Electrochemical control.* In an electrochemical setup where the friction could be measured between the conducting surfaces, one can independently control the potential of each electrode relative to a reference electrode in the bulk of electrolyte. Although, we did not perform simulations at constant potentials, based on the results presented above, it is straightforward to foresee the corresponding *voltage effect on friction.* The voltage will induce a shift of the slippage plane form the solid-liquid interface to the interior of the film. With the increase of the potential of each surface, the friction would grow to a maximum and then decrease.

We expect *asymmetric electrostriction / electroswelling of the film*: (i) If we polarize both surfaces negatively, we first slightly decrease the thickness of the film, and then increase it to a plateau. (ii) If we polarize both surfaces positively, the film would get thinner, with the cations first leaving the surface, and then majorly leaving the confined region. (iii) If the electrode potential changes for one of the surfaces, the trends would be the same, but the magnitude of the effect would be roughly two times smaller. (iv) If we change the potentials of the electrodes in the same direction but in an unequal manner, the trends will be the same, except that the dip at medium negative potentials will be smeared. (v) For oppositely charged electrodes the interplay between the mentioned effects can be more complicated.

In the following we list the generic trends for friction between charged surfaces lubricated with RTIL nano-films, as can be inferred from our work:

1. The friction force reaches a maximum as a function of surface charge. This maximum corresponds to the charge at which the slippage plane shifts one-layer-away from the solid-liquid interface to the boundary between the counterion- and coion-rich layers.
2. The nanoscale film will exhibit the electrostriction / electroswelling effect, which will be asymmetric with respect to the sign of the surface charge, as long as the cations and anions have different molecular shapes, chemical structure and intramolecular charge distribution. We have discussed above such effect for BMIM PF6.

All these predictions await experimental verification. On the other hand, this is just the beginning, and a set of challenging investigations is ahead:

- Although general trends have been summarized both for controlled surface charge and surface electrical potential, the continuous charge variation through the voltage control [26] deserves a careful study.
- Even the so-called hydrophobic RTILs adsorb water. The presence, and the debated location of electrosorbed water, within the lubricating RTIL nanofilms [27,28] is of crucial importance for friction [29,30].
- The possible effect of image charges remains to be investigated. It will be different for mica surfaces, metallic surfaces, or graphene layers on dielectric substrates. Account for image forces may change the distribution of ions and water across the film.

Further studies along those lines are needed for rational applications of electrotunable lubricity with RTILs.

**Computational Methods.**

Non-equilibrium Molecular dynamics simulations of friction were performed using Grand Canonical Molecular Dynamics (GCMD) simulations.[31,32,33] In this method the confined region is simulated explicitly by setting two solid plates in the bulk fluid (see Fig. 1). The confined region is in contact with a reservoir fluid, hence the chemical potential in the confined region is determined by the chemical potential of the reservoir by allowing the ionic liquid to enter or leave that region as a response to changes in the normal pressure, temperature or surface charge. It has been shown that the GCMD method preserves the chemical potential along the simulation box, in confined and non-confined regions.[33] The atoms in the walls interacted through a strong Lennard-Jones potential with the interaction strength, 50 kJ/mol, while the Lennard-Jones interaction between atoms in different walls was set to 1 kJ/mol. The effective Lennard-Jones diameter of the atoms in the wall was set to 0.3218 nm. A typical wall consisted of 2838 atoms. The lattice parameter of the solid blocks was set to 0.36nm in a fcc lattice. The slabs were arranged in the simulation box on the *xy* plane, and they were periodic in the *y* direction (see Figure 1).

Our simulations involved typically 6,000 cations and anion pairs that were added to the simulation box containing the solid plates (see Fig. 1). In addition to the Coulombic interactions, the beads interacted with other beads or atoms in the solid plates according to short range Lennard-Jones interactions with the parameters [17]: $(\sigma, \varepsilon)_{C1}$= (0.504 nm, 1.83 kJ/mol), $(\sigma, \varepsilon)_{C2}$=(0.438 nm, 2.56 kJ/mol) and $(\sigma, \varepsilon)_{C3}$= (0.341 nm, 0.36 kJ/mol) and $(\sigma, \varepsilon)_{anion}$= (0.564 nm, 4.71 kJ/mol) for the anions. The following charges, $q_{C1} = 0.1848e, q_{C2} = 0.4374e, q_{C3} = 0.1578e, q_{anion} = -0.78e$ have been assigned to the beads, C1, C2, C3 of the cation, and to the anion, respectively. Further we used $(\sigma, \varepsilon)_{is}$=(0.3218 nm, 22.5 kJ/mol) for the ion-solid plates interactions. The cross interactions between sites of type *i* and *j* were constructed from these parameters using standard combining rules, *i.e.*, $\sigma_{ij}=(\sigma_i + \sigma_j)/2$ and $\varepsilon_{ij} = \sqrt{\varepsilon_i \varepsilon_j}$. We have also performed simulations for a several values of ion-wall interactions. We did not find differences in the main conclusions discussed in the paper.

In order to achieve well-equilibrated configurations with the appropriate number of confined layers (wall-wall separation) we first performed simulations with applied normal pressure but no shear. These simulations involved about ~8 ns. After this pre-equilibration the system was subjected to shear for about 16 ns, by moving the mobile solid plate at constant velocity as discussed in the text. Shearing heats the entire system. To prevent this we coupled the solid plates to a thermostat (using the v-rescale method [34]) at the 450 K. This served to maintain the temperature of the ions, which were not coupled to an explicit thermostat. All the simulations were performed by coupling the entire system to an anisotropic Berendsen barostat set at 1 bar pressure, which was applied in the *x* and *z* directions, keeping constant the box length in the direction of pulling, *y*. Periodic boundary conditions were imposed in all directions.

The equations of motion were integrated with the Leap-Frog algorithm and a time step of 2 fs, using the code Gromacs v. 4.6.3.[35] The long range interactions arising from the Coulombic charges were computed using the 3D –Particle Mesh Ewald summation using "tin-foil" boundary conditions. Following our previous work,[8] we screened all the charges in the system by using an effective dielectric constant of 2, in order to account for the electronic polarizability of the liquid. Configurations were stored every 6 ps for further structural and dynamics analyses.

## ASSOCIATED CONTENT

Supporting Information

Results of simulations of friction with weaker ion-mica Lennard-Jones interactions, simulations of wetting and supplementary figures. The Supporting Information is available free of charge on the ACS Publications website at

## AUTHOR INFORMATION

Corresponding Authors

*E-mails: yofar008@gmal.com ; f.bresme@imperial.ac.uk ; a.kornyshev@imperial.ac.uk;

urbakh@post.tau.ac.i

Notes

The authors declare no competing financial interest.

**ACKNOWLEDGEMENT**

FB acknowledges the financial support from EPSRC (EP/J003859/1) and the award of an EPSRC Leadership Fellowship. M. U. acknowledges the financial support of the Israel Science Foundation, Grant No. 1316/13. AAK and MU thank British Council for the support of MU visit to Imperial where the work has been accomplished.